## Article title
*Vacuum cleaving of superconducting niobium tips to optimize noise filtering and with adjustable gap size for scanning tunneling microscopy*

## Authors

*Carolina A. Marques[1, *], Aleš Cahlík[1], Berk Zengin[1], Tohru Kurosawa[2], Fabian D. Natterer[1, *]*

## Affiliations
1. Department of Physics, University of Zurich, Winterthurerstrasse 190, CH-8057 Zurich, Switzerland
2. Department of Applied Sciences, Muroran Institute of Technology, Muroran 050-8585, Japan

## Corresponding author's email address
carolina.dealmeidamarques@physik.uzh.ch, fabian.natterer@physik.uzh.ch

## Keywords

## Abstract

*Superconducting (SC) tips for scanning tunneling microscopy (STM) can enhance a wide range of surface science studies because they offer exquisite energy resolution, allow the study of Josephson tunneling, or provide spatial contrast based on the local interaction of the SC tip with the sample. The appeal of a SC tip is also practical. An SC gap can be used to characterize and optimize the noise of a low-temperature apparatus. Unlike typical samples, SC tips can be made with less ordered materials, such as from SC polycrystalline wires or by coating a normal metal tip with a superconductor. Those recipes either require additional laboratory infrastructure or are carried out in ambient conditions, leaving an oxidized tip behind. Here, we revisit the vacuum cleaving of an Nb wire to prepare fully gapped tips in an accessible one-step procedure. To show their utility, we measure the SC gap of Nb on Au(111) to determine the base temperature of our microscope and to optimize its RF filtering. The deliberate coating of the Nb tip with Au fully suppresses the SC gap and we show how sputtering with $Ar^+$ ions can be used to gradually recover the gap, promising tunability for tailored SC gaps sizes.*

- Oxide free superconducting STM tips
- RF filter optimization

## Graphical abstract

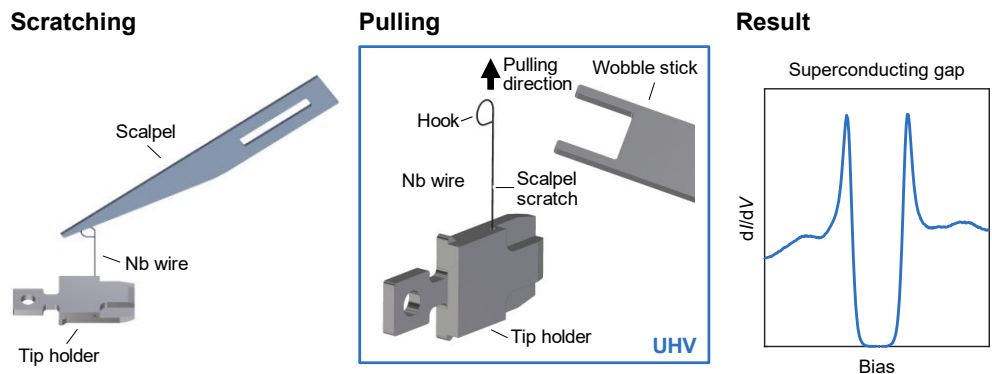

**Method details**

Scanning tunneling microscopy (STM) allows the study of atomically sharp tunneling junctions at sub-Kelvin temperatures and under large magnetic fields. A versatile feature of STM is that one can choose tip materials with desired properties, including materials for spin-polarized and superconducting (SC) tips. The use of SC tips in STM provides enhanced energy resolution(3), spin-polarized measurements(4, 5), enables the study of Yu-Shiba-Rusinov states(6), and the creation of Josephson tunneling junctions(7, 8). Additionally, SC tips can be used to optimize the noise performance in an STM, by using the gap function as a monitor for the presence of radiofrequency (RF) noise in the tunneling junction and thereby cables, as well as to determine the effective temperature of the STM junction(9–11).

Superconducting (SC) tips are frequently prepared by dipping a metallic tip (Pt/Ir or W) into a SC single crystal such as Nb and Pb to encapsulate the normal metal apex with the superconductor(3, 12). This requires, in addition to Nb or Pb samples, surface preparation tools, such as sputter guns and heating stages. Alternatively, a SC tip can also be directly made from SC wires, either by electrochemical etching(13) or mechanical cleaving(1, 2). The advantage of mechanical cleaving in vacuum is that no additional treatment steps to remove the natural oxide layer are necessary, which would consist of annealing, field-emission, or sputtering in the vacuum system. Here we use a one-step method to prepare pristine Nb tips directly in vacuum, following the procedure reported in Ref.(2). For demonstration, we use a commercial low-temperature scanning tunneling microscope (STM, Createc Fischer GmbH). This in-situ method can immediately provide a SC tip without additional treatment steps. We verify the utility of this approach by measuring the SC gap, Δ, in tunneling spectroscopy, using the Nb tips to efficiently optimize the RF filtering, and determine the effective base temperature of our STM. We furthermore coat the tip with Au by identing it into an Au(111) single crystal and are able to recover the Nb gap after sputtering the tip with $Ar^+$ ions. The gradual recovery of the SC gap with sputtering time, would allow to deliberately prepare tips with a tunable gap size in the range [0, Δ].

The procedure to obtain SC tips is illustrated in Figure 1a and works as follows: We take a Nb wire of thickness 0.125 mm (purity 99.9%, Goodfellow) and we cut it to a length of about 2 cm. On the far end, we make a loop with a knot (Figure 1b). We insert the wire onto a tip holder and we take a sharp blade to make a slight notch at an angle of about 45º into the malleable Nb wire, 3 mm away from the edge of the tip holder (Figure 1a, c). This distance will define the effective tip-length and would have to be adjusted according to the dimensions of a specific setup. We then glue the wire to the tip holder (Figure 1a) using conductive silver epoxy (EPOTEK 20E). We use the same glue on the knot of the loop to secure it while pulling (Figure 1b) and cure the glue for 90 min at 150ºC. The notch creates a weak point in the wire (Figure 1b, left inset), where it will break when the loop is pulled. To demonstrate the cleaving concept, we pulled several tips apart in ambient conditions and investigated the resulting apices to develop a practice for the later in-situ cleave. The optical microscope image in the right inset of Figure 1c shows the result of a test-cleave after pulling at the wire loop.

An example of a Nb wire with loop and notch glued to a tip holder can be seen in Figure 1d. We transfer it via the loadlock and preparation chamber to our STM chamber kept at a base pressure better than $3\times10^{-10}$ mbar. There, we use the wobble stick to pull it apart at the loop (Figure 1e), indeed breaking the Nb wire at the notch (Figure 1f). As can be seen in Figure 1f, the tip is now ready for a transfer into the STM head.

As with other tips, we use the reflection of the Nb tip in the polished sample surface for the coarse alignment on Au(111) (Figure 1g). Despite the clearly tilted wire, the Nb tip allows imaging of single impurities and of the standing wave patterns of the scattered Au Shockley surface(14) right upon the first approach (Figure 2a). This clearly demonstrates the immediate utility of the Nb tip for regular surface science experiments. In addition, the differential conductance (Figure 2b) shows the expected superconducting gap Δ of pure Nb. The gap function is characterized by absent zero-bias conductance (fully gapped) and coherence peaks at ±Δ. Accordingly, the fitting to a Dynes function(15) with a quasiparticle-lifetime broadening constant of 0.05 µeV is shown in Figure 2b, giving a gap of Δ = (1.25 ±0.01) mV and an STM junction temperature of T = (1.21 ± 0.02) K.

The SC gap of the Nb tip can also serve as an accessible way to optimize RF filtering and determine the effective energy resolution of an STM setup(9–11). Our experience offers a good example: The blue trace in Figure 3a shows our setup after RC low pass filtering bias, current, thermometry, and piezo scanner lines prior to using an SC tip for optimization. Although this filtering improved the gap shape, we still had finite zero-bias conductance, with broadened coherence peaks slightly shifted towards higher energies. After the addition of pi-filters (Tusonix 2499-003-X7R0-101MLF plus 10 kΩ resistor) on the piezo scanners and disconnecting the coarse motion walkers, we were finally able to detect a fully formed SC gap and sharp coherence peaks (black line). The width of the coherence peaks is now only limited by thermal broadening instead of RF noise. Here, the SC gap provided by the cleaved Nb wire clearly helped us to rapidly identify and mend problematic wiring.

To determine whether it is possible to recover the Nb apex after crashing into the sample, we first dip the tip a few dozen nm into the Au crystal until the zero bias conductance matches the quasiparticle conductance (Figure 3b, black trace). Note that at this point, we are unable to recover the gap by voltage pulsing to 10 V, nor did the gap recover when going into a mild field emission regime at 30 V. To remove the Au accumulated at the apex, we sputter the tip with Ar$^+$ ions (3 keV, 10 µA/cm$^2$). During sputtering, we moved the tip about 4 mm off center of the Ar$^+$ beam, resulting in the removal of Au in the nm range. After a cycle of 45 min, a SC gap reappears albeit with a smaller width and larger zero-bias conductance. This is consistent with the proximity effect between a SC and a normal metal(16). We sputtered the tip three more times for a total of 192 min = (45 + 45 + 47 + 55) min, each time verifying the progression of the SC gap (Figure 3a), until we recover a superconducting gap without zero-bias conductance. More sputtering cycles are needed to fully recover the as-cleaved fully gapped state. The gradual recovery of the SC gap allows the deliberate tuning of the gap size by an appropriate sputtering time, which could be useful in Josephson junction geometries to match the SC gap size of the sample ($\Delta_{tip} = \Delta_{sample}$)(17).

To test reproducibility of the tip-cleaving method, we prepared a total of 9 Nb wires. While 7 cleaves showed tips with a superconducting gap, two tips were insulating. Out of the 7 SC tips, four showed a fully gapped spectrum with no zero-bias conductance at 0 mV, whereas three show a smaller gap with finite conductance. Examples of tunneling spectra for different tips after cleaving and first approach on Au(111) are shown in Figure 3c.

In summary, the vacuum cleaving of Nb-wires allows for a quick and easy way to prepare superconducting tips for STM experiments. We have shown that this method is very forgiving; it generates sharp superconducting tips with fair yield, even before ideal experimental execution. This method is an accessible way to generate a well characterized fully gapped superconducting spectrum in an ultra-high vacuum (UHV) STM setup. We showed that these tips can be used to optimize the RF filtering and energy resolution of an STM. By coating the SC tip with Au, we demonstrated how the gap size can be tuned to the experimental requirements by taking advantage of the proximity effect and by gradually changing the size of the Au cluster with Ar$^+$ sputtering.

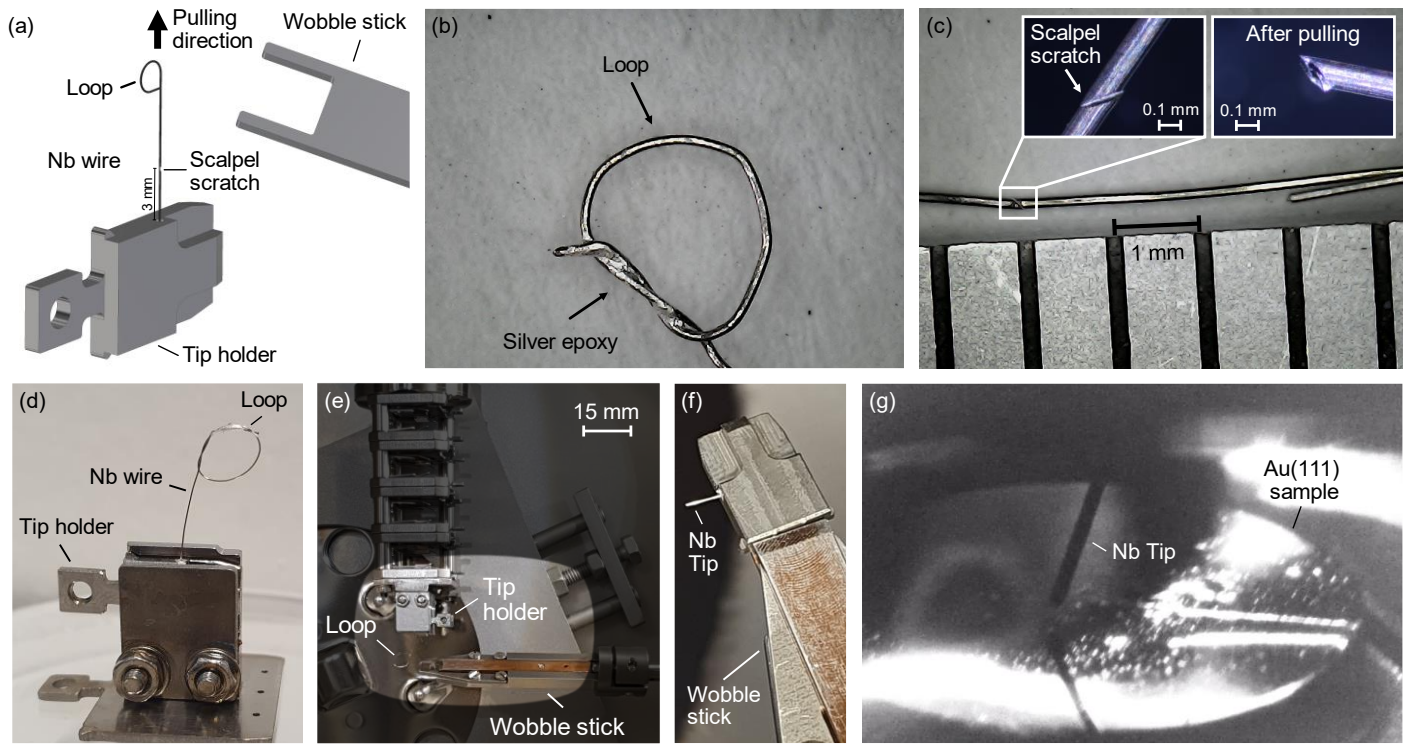

*Figure 1: In-situ cleaving of Nb tips.* (a) Sketch of the Nb tip preparation setup. A pre-notched Nb wire is glued to the tip holder. At the far end, the Nb wire is wound into a loop which allows pulling the wire apart with the tongs of a wobble stick and break it at the notch. (b) Photography of the loop that is secured with silver epoxy to ensure that it withstands the pulling. (c) Photography of a Nb wire weakened with a scalpel. The left inset is an optical microscope image of the scalpel scratch and the inset on the right shows an optical microscope image of the Nb wire after pulling it apart in ambient conditions. (d) Photography of the Nb wire with loop glued to the tip holder and with a scalpel scratch ready to load into the load-lock chamber. (e) Image still of the actual in-situ cleaving of the Nb tip inside the vacuum chamber. The tip holder is placed in one of the tip storage slots. The wobble stick is shown reaching for the loop with one of its tongs. (f) Photography of the Nb tip right after cleaving. (g) Photography of the aligned Nb tip atop an Au(111) single crystal inside of the STM head.

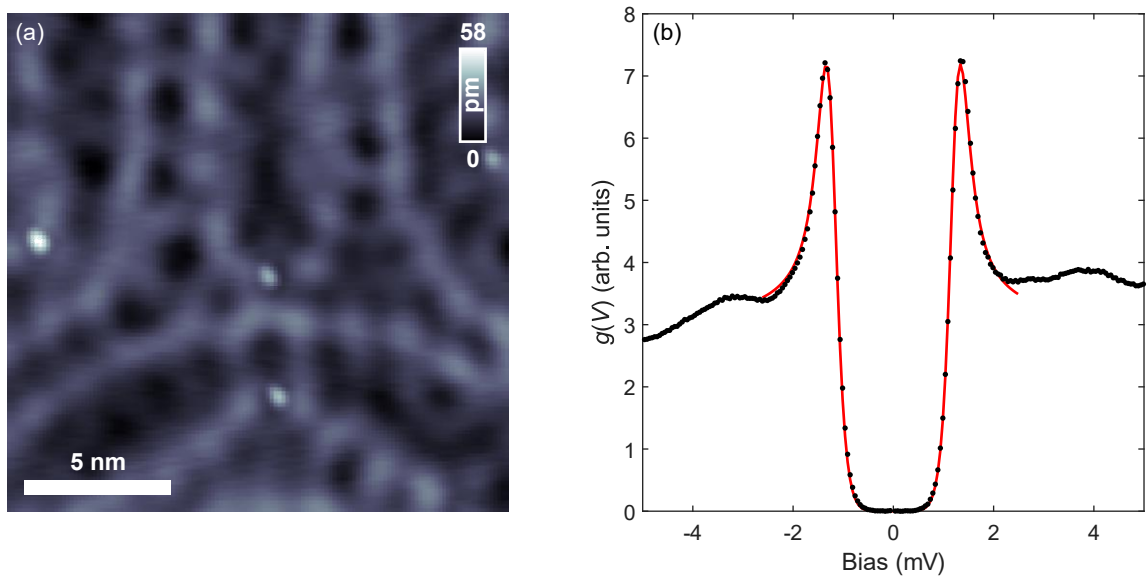

*Figure 2: Superconducting tip characterization.* (a) Topography of Au(111) surface measured with as-cleaved Nb tip showing the Shockley surface state as standing wave ripples emerging from impurities. The latter (white dots) prove the sharpness of the Nb tip upon approach, without requiring any additional tip treatment ($V_S$ = 10 mV and $I_S$ = 1.2 nA). (b) Differential tunneling conductance $g(V)$ showing the superconducting gap expected of pristine Nb ($V_S$ = 10 mV, $I_S$ = 1.2 nA, $f_{mod}$=613 Hz, $V_{mod}$ = 50 µV). The red line shows the fit to a Dynes function with a superconducting gap of Δ = (1.25 ±0.01) mV that yields a critical temperature of $T_c = \frac{\Delta}{1.76 k_B}$ = 8.2 K. The fit furthermore establishes our system temperature of T = (1.21±0.02) K.

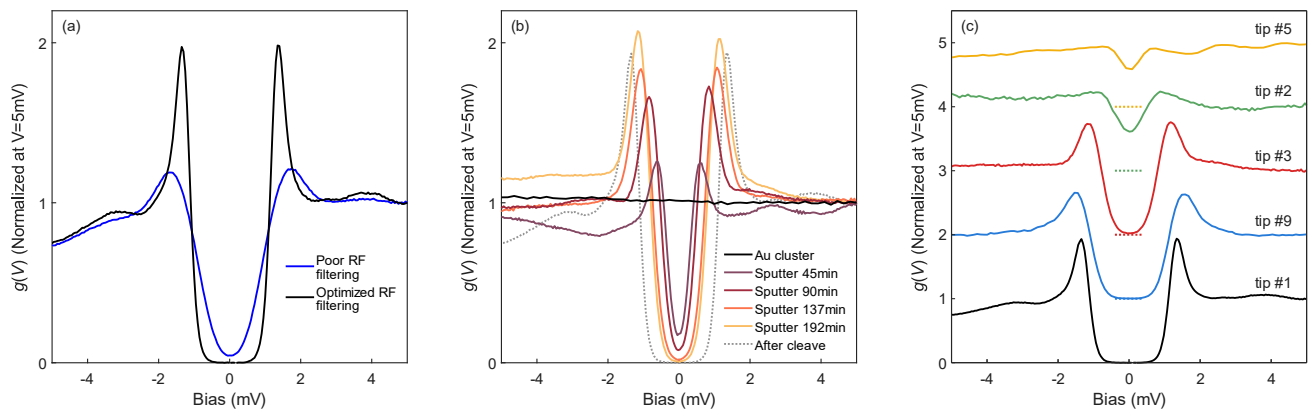

*Figure 3: Utility of Nb tips for RF filtering optimization, adjustable gap size, and cleaving reproducibility.* (a) Nb tips can also help optimize RF filtering of an STM as the gap function is sensitive to the presence of RF noise. Differential conductance spectra before (blue) and after (black) optimizing the RF filtering of the STM. ($V_s$ = 10 mV, $I_s$= 1.2 nA, $f_{mod}$=613 Hz, $V_{mod}$ = 50 μ (b) Recovering Nb tip superconductivity after deliberate dipping the apex several dozen nanometers into the Au crystal. The differential tunneling conductance spectra show how we progressively recover superconductivity from a normal state Nb tip (coated by Au, black) to a superconducting gap without zero conductance by $Ar^+$ ion sputtering cycles, effectively enabling a tuning of the gap size. (c) Set of tunneling spectra that show the resulting superconducting gaps of different cleaves. The curves are displaced vertically for better visualization and the dashed lines indicate the g(V=0mV) for each subsequent curve.

## CRediT author statement
*__Carolina A. Marques:__ Conceptualization, Methodology, Investigation, Visualization, Validation, Writing - Original Draft, Funding acquisition, __Aleš Cahlík__: Methodology, Validation, Investigation, __Berk Zengin__: Methodology, Investigation, __Tohru Kurosawa__ : Methodology, __Fabian D. Natterer:__ Conceptualization, Methodology, Writing - Original Draft, Supervision, Funding acquisition.*

## Acknowledgments
*Funding: This work was supported by the Swiss National Science Foundation [PP00P2_211014, 200021_200639] and through the Federal Commission for Scholarships for Foreign Students for the Swiss Government Excellence Scholarship (ESKAS No. 2023.0017) for the academic year 2023-24.*